\documentclass[conference]{IEEEtran}
\IEEEoverridecommandlockouts
\usepackage{amsmath,amssymb,amsfonts}
\usepackage{algorithmic}
\usepackage{graphicx}
\usepackage{textcomp}
\usepackage{xcolor}
\usepackage{multirow}
\usepackage[acronym]{glossaries}
\usepackage{hyperref}

\newacronym{ADM}{ADM}{Automated Decision-Making}
\newacronym{TPR}{TPR}{true positive rate}
\newacronym{FPR}{FPR}{false positive rate}
\newacronym{PPV}{PPV}{positive predictive value}
\newacronym{FOR}{FOR}{false omission rate}

\begin{document}

\title{A Systematic Approach to Group Fairness\\in Automated Decision Making
}

\author{\IEEEauthorblockN{Corinna Hertweck}
\IEEEauthorblockA{\textit{Zurich University of Applied Sciences,} \\
\textit{University of Zurich}\\
Zurich, Switzerland \\
corinna.hertweck@zhaw.ch}
\and
\IEEEauthorblockN{Christoph Heitz}
\IEEEauthorblockA{\textit{Zurich University of Applied Sciences} \\
Winterthur, Switzerland \\
christoph.heitz@zhaw.ch}
}

\maketitle

\begin{abstract}
While the field of algorithmic fairness has brought forth many ways to measure and improve the fairness of machine learning models, these findings are still not widely used in practice.
We suspect that one reason for this is that the field of algorithmic fairness came up with a lot of definitions of fairness, which are difficult to navigate.
The goal of this paper is to provide data scientists with an accessible introduction to group fairness metrics and to give some insight into the philosophical reasoning for caring about these metrics.
We will do this by considering in which sense socio-demographic groups are compared for making a statement on fairness.
\end{abstract}

\section{Introduction}

The field of algorithmic fairness is, to a large extent, concerned with the fairness of \gls{ADM} systems.
\gls{ADM} refers to the process of making decisions about individuals in an automated way based on the data that is available about these individuals.
Examples are the automated approval of credit loan applications based on, e.g., the credit loan history and income of the applicant, the automated acceptance to universities based on grades and standardized test scores or the automated invitation of applicants to job interviews based on their CV.
\gls{ADM} typically involves a prediction $\hat{Y}$ of a non-observable feature $Y$ of individuals.
An example of this would be the prediction $\hat{Y}$ of whether or not a credit loan applicant would repay the loan they applied for.
The automated decision is then based on this prediction.
In many instances, it has been shown that such prediction-based decision systems lead to unintended unfairness and discrimination (see, e.g., \cite{machine-bias, obermeyer2019dissecting} for specific cases of unintended algorithmic discrimination or see \cite{orwat2019diskriminierungsrisiken} for an overview of the topic).
These instances have made clear that the fairness properties of such decision systems have to be studied since fairness is not achieved automatically, but has to be consciously implemented.

Our paper provides data scientists with an entry point to the many metrics that fall into the category of \textit{group fairness}.
Measures in this category evaluate \gls{ADM} system's decisions with respect to different socio-demographic groups (e.g., different gender or racial groups).
The basic question is: \textit{Are there systematic differences in the decisions for the groups?}
When evaluating fairness through group fairness metrics, we thus always compare groups and look for differences in treatment.
We analyze the metrics not by focusing on their mathematical definitions, but by looking at how exactly the decisions for socio-demographic groups are compared to each other by the given fairness metrics as this provides a basis for a moral interpretation.
To help data scientists answer the question "What fairness measures should I care about, and why?", we will provide a short (and necessarily incomplete) introduction to the philosophical arguments that have been made about the discussed fairness metrics.
\section{Related work}

The fairness discussion is not a new discussion that appeared with the increased usage of \gls{ADM} systems.
Quite the opposite is true: Questions of fairness and justice have been discussed for thousands of years, with a long history in philosophy, sociology, law etc. However, \gls{ADM} systems require quantitative measures of fairness, and the algorithmic fairness literature of the last years has brought forward many different such measures:
In \cite{narayanan2018translation}, Arvind Narayanan discusses 21 fairness metrics, but notes that there are many more. The fairness auditing tool AIF360 claims to "[include] over 71 bias detection metrics" \cite[p. 2]{aif360-oct-2018}.
However, it can be difficult to understand the specifics of these different fairness measures that have been proposed in the literature and are implemented in state-of-the-art tools.
Some publications try to categorize fairness measures (see, e.g., \cite{verma2018fairness, mitchell2021algorithmic}), but these works usually lack intuitive explanations of the moral reasons to enforce the discussed fairness measures and instead focus on their mathematical definitions.
This may create confusion on the side of practitioners about how to implement fairness in their data-based decision algorithms.

There is already a considerable amount of literature that discusses different fairness measures and in particular group fairness measures, which we focus on in this paper.
Since most group fairness metrics are calculated from the values found in confusion matrices, existing literature oftentimes approaches the introduction of different group fairness measures from this angle:
They show how these measures can be calculated from the confusion matrix (see, e.g., \cite{verma2018fairness, mitchell2021algorithmic, kim2020fact, miron2020addressing, hellman2020measuring, makhlouf2020applicability}).
Recent research (see \cite{nyarko2020breaking, goel2020predictions}), however,  experimentally showed that confusion matrices are not intuitive and that it takes time to fully understand them.
From this, we conclude that it is hard to grasp the moral meaning of fairness metrics based on confusion matrices.
Another explanatory approach is chosen in \cite{fairmlbook}: Here, the authors define common group fairness measures as (conditional) independence statements.
While this approach is formally very elegant, it is difficult to interpret with respect to the concept of fairness and justice that is represented by each metric.

Our paper attempts to introduce group fairness metrics in a more intuitive way by showing that each metric corresponds to an analysis of the decisions made about a specific part ("subpopulation") of the population on which the system operates.
For example, we might consider all members of the population, or only those who are negatively affected by the decision algorithm.
We argue that it is exactly this choice of the subpopulation which establishes which theory of fairness a specific metric aligns with. 
\section{What is a "fair \gls{ADM} system"?}\label{sec:what-is-fairness}

The question of what constitutes fairness is a heavily debated philosophical question.
In this section, we will first show that the intuitive interpretation of fairness as non-discrimination cannot be easily applied to the field of algorithmic fairness.
We then introduce another philosophical concept, egalitarianism, which can be applied to understand when \gls{ADM} systems are considered to be unfair.

\subsection{Non-discrimination}

Often, fairness is equated with non-discrimination, which is legally defined in anti-discrimination laws.
Article 8 of the Swiss federal constitution, for example, states that "no person may be discriminated against, in particular on grounds of origin, race, gender, age, language, social position, way of life, religious, ideological, or political convictions, or because of a physical, mental or psychological disability."
The law thus defines several properties on which discrimination is specifically prohibited.
We will refer to these as \textit{sensitive attributes} (e.g. origin, race and gender).
While equating fairness with non-discrimination seems intuitive, it turns out that the \textit{philosophical} concept of discrimination is actually not applicable to "unfair" machine learning (see \cite{binns2018fairness} for a detailed analysis).
One of the reasons that Reuben Binns \cite{binns2018fairness} gives for this is that one may see discrimination as morally wrong only when it is intentional.
However, it is difficult to speak of intention in the case of \gls{ADM} systems which we would typically not portray as conscious, autonomous actors.
The questions that this finding leaves us with are: What is fairness in \gls{ADM} systems if the philosophical concept of non-discrimination cannot be applied? On what philosophical grounds should an \gls{ADM} system be assessed?

\subsection{Egalitarianism}

A philosophical concept that might better help us understand when \gls{ADM} systems are unfair is \textit{egalitarianism}.
Binns describes egalitarianism as "the idea that people should be treated equally, and (sometimes) that certain valuable things should be equally distributed" \cite[p. 5]{binns2018fairness}.
As he further argues, this does not prohibit inequalities.
Distributive inequalities might sometimes be justified and considered to be fair.
What fair inequalities are cannot be objectively determined though -- instead it depends on the philosophical theory of justice one holds.
One may, for example, argue that differences between people are only just if they are not caused by brute luck, such as being born into a wealthy family (this view is known as luck egalitarianism and defended by philosophers such as Ronald Dworkin and G. A. Cohen, see, e.g., \cite{dworkin_what_1981, cohen_currency_1989}).
On the other side, one could claim that such inequalities are fair under certain circumstances (a definition of such circumstances is given and defended by the American philosopher John Rawls \cite{rawls2001justice}).
To summarize, fairness in \gls{ADM} systems can be understood as some form of equal treatment or equal distribution between people.
People, however, disagree about what constitutes equal treatment, what justifies inequalities and how this should be measured -- which is how we end up with the many fairness metrics we find in the literature \cite{jacobs2021measurement}.
\section{Measuring fairness}

Data-based decision algorithms typically operate in a context where only partial information is available at the time of decision making: The missing information $Y$ is not available and thus replaced by a prediction $\hat Y$, which is then used to take a decision.
However, this means that decisions might be wrong on the individual level: a bank might give a loan to someone who does not pay this loan back; a student might be accepted to a college even though they do not succeed in their studies there.
When assessing the fairness of a decision system, we may thus not be interested in random mistakes, but in systematic patterns in the decision.
In other words: We may call a decision algorithm unfair if it \textit{systematically} disadvantages one group with respect to the other. This concept is known as {\em group fairness}.

Since the decisions in the end still affect individuals, both the philosophical theories of justice as well as the law tend to discuss justice with respect to individuals.
The algorithmic fairness literature actually provides approaches for measuring fairness on an individual level.
The concept of \emph{individual fairness}, for example, asks whether "similar" individuals receive "similar" decisions (it is up to the decision-maker to define what individuals and what decisions are considered to be "similar") \cite{dwork2012fairness}.
\emph{Counterfactual fairness} asks whether individuals would have received the same decision if they were (born as) a member of another socio-demographic group \cite{kusner2017counterfactual}.

In this paper, we focus on group fairness measures which assess whether a certain property of the decision algorithm, averaged over the members within different socio-demographic groups, is equal across these groups.
The averaging reveals the systematic effects of the decision system. 
In the context of this paper, we therefore define fairness as socio-demographic groups having to be equal with respect to a certain property of the \gls{ADM} system.
\section{Understanding group fairness measures}

Group fairness in the context of data-based decision making means comparing how different socio-demographic groups are treated by the \gls{ADM} system.
In this section, we analyze how exactly socio-demographic groups can be compared in the fairness-context.
As it turns out, there are different options, and each one constitutes its own "fairness metric" (or "fairness measure").
We also note that there are similarities between some metrics which lead to the broader category of "fairness criteria" \cite{fairmlbook}.

Table \ref{tab:math} provides an overview of the fairness metrics we discuss in this section.
Each fairness metric that we consider is based on three elements:
\begin{itemize}
    \item The definition of socio-demographic groups to be compared: The choice of which socio-demographic groups are compared is crucial. By, for example, comparing only women and men, unfair treatment of non-binary people cannot be detected. By comparing different gender groups and, in a separate evaluation, different racial groups, we might not notice that black women are particularly harmed by a system \cite{buolamwini2018gender}.
    \item The relevant subpopulation on which those groups are compared: This could, for example, be the subpopulation of people negatively affected by the decision.
    \item The relevant comparison property: The definition of {\em what} should be equal when comparing these socio-demographic groups.
\end{itemize}
We will see that the metrics differ in the second and third element, which provides an intuitive, hands-on understanding of these metrics.
This understanding of the metrics helps us in taking a step towards answering the arguably biggest question in implementing group fairness metrics: Which metric should one aim to fulfill?

Our discussion of the fairness metrics will be based on the example of credit lending.
We assume that a machine learning model predicts whether or not an applicant pays back their loan ($\hat Y$).
The model has been trained on a training set consisting of the sensitive attribute $A$, which in our case is the gender of the applicant (to simplify the explanation, we assume that we only compare female to male applicants), additional attributes $\vec X$, such as their credit history and income, and the binary indicator $Y$ of whether or not the applicant repaid their loan.
A subsequent decision algorithm then takes the decision $D$ (whether a given applicant should receive a loan) based on the prediction $\hat Y$ (calculated from the features $\vec X$ and $A$), combined with business related parameters such as current interest rates.\footnote{
A large part of the algorithmic fairness literature is focused on cases where $D$, $Y$ and $A$ are binary. In the interest of making this paper as accessible as possible, we will follow this convention and do not discuss more complex settings.
}
This decision model is tested with a representative pool of credit loan applicants consisting of women and men by checking for systematic differences in the decision pattern.
Hereinafter, we will refer to this testing data as the "full population" and to parts of it as "subpopulations".

The most obvious way to compare women and men is to look at the decision outcomes for the full population, i.e., to compare the decisions for \emph{all men of the population} to the decisions for \emph{all women of the population}.
This leads to the fairness criterion called "independence" (see \cite{fairmlbook} and Table \ref{tab:math}), which is also known as "statistical parity" or "demographic parity" in the case of classification.
It answers the question: Are women and men equally likely to get a loan?\footnote{
Alternatively, we might ask whether women and men are equally likely to be denied their credit loan application. In the case of statistical parity, equality (inequality) in the acceptance rate implies equality (inequality) in the rejection rate. Thus, assessing the rejection rate does not provide new information about the fairness of the model -- it just frames it in a different way. This is also the case with the following metrics for which we compare the credit approval rate. For better readability, we restrict the discussion to the approval rate.
}
Note that this does not require equal numbers of women and men to be accepted, but equal shares of women and men to be accepted.

Instead of considering the full population, we might also restrict the fairness analysis to subpopulations.
For example, we might ask for the probability of getting a loan in the subpopulation of applicants with a stellar credit history and high income ($\rightarrow$ "conditional statistical parity") or in the subpopulation of applicants who actually repay their loan (i.e., "good debtors" $\rightarrow$ "true positive rate").
For a systematic overview of the possible subpopulations, we recall the basic structure of the decision problem, which is characterized by the vector of variables $(A,\vec X,Y,D)$.
While the variable $A$ defines the groups that are being compared (in our case: men and women), all other variables might be used to define subpopulations.

First, we might restrict the analysis to a subpopulation characterized by a specific range of $\vec X$.
This leads to "conditional statistical parity" (see Table \ref{tab:math}).\footnote{
Even though conditional statistical parity is only a relaxation of statistical parity \cite{fairmlbook} and as such might not demand another detailed explanation, we add it to this list as it seems to hold intuitive value for many people. We see this, for example, in discussions about the gender pay gap \cite{grimshaw2002adjusted}.
} 

Another option is to restrict the fairness analysis to individuals according to their value of $Y$, e.g., to individuals with $Y=1$.
These individuals still differ in their value of $D$, i.e. in whether their loan application is approved or not, and a natural approach to check for group differences is to compare the acceptance rate of men to that of women. However, this is done only within the subpopulation of individuals actually repaying their loan -- it thus represents a fairness assessment on the group of "good debtors".
In the case of binary classifiers, the combination of both resulting fairness metrics (for the subpopulations of either $Y=0$ or $Y=1$) is called "separation" \cite{fairmlbook} (see Table \ref{tab:math}).

Finally, we might also restrict the analysis to individuals with the same decision $D$, e.g. to individuals with $D=1$.
In this case, assessing the average approval rate makes no sense.
What can be evaluated though is the average value of $Y$ for this subpopulation, i.e., the average repayment rate of people with $D=1$. Equal repayment rates among accepted female and male credit applicants might be seen as a proof that the decision does not lead to systematic bias.
The combination of both resulting fairness metrics (for the subpopulations of either positive or negative decisions) is called "sufficiency" \cite{fairmlbook} (see Table \ref{tab:math}).

The important realization here is that there is a limited number of ways to compare socio-demographic groups (here, men and women) in the case of the decision problem with the four variables $A,\vec X, Y,D$.
In fact, most of the group fairness measures proposed in the literature fall into one of the three categories of either independence, separation, or sufficiency \cite{fairmlbook}.\footnote{Note that, formally, one might also assess the average $Y$ (instead of the average $D$) in the case of the full population.
However, this metric does not reflect a property of the decision algorithm, but rather a statistical property of the population and is typically referred to as the "base rate".
In our example, it is the rate at which women and men in the dataset repay their credit loans.
}

\begin{table*}[htbp]
\caption{Structured overview of fairness metrics}
\scriptsize
\label{tab:math}
\begin{tabular}{|l|l|l|l|l|}
\hline
\textbf{Fairness criterion}           & \textbf{Fairness metric}             & \textbf{Subpopulation} & \textbf{Property} & \textbf{Equation}                            \\ \hline
\textbf{Independence}                 & Statistical parity                   & -                      & $P(D=1)$          & $P(D=1|A=a)=P(D=1|A \neq a)$                       \\ \hline
\textbf{Relaxation of independence}   & Conditional statistical parity             & $\vec{X}=x$            & $P(D=1)$          & $P(D=1|\vec{X}=x, A=a)=P(D=1|\vec{X}=x, A \neq a)$ \\ \hline
\multirow{2}{*}{\textbf{Separation}}  & Parity of true positive rates        & $Y=1$                  & $P(D=1)$          & $P(D=1|Y=1, A=a)=P(D=1|Y=1, A \neq a)$             \\ \cline{2-5} 
                                      & Parity of false positive rates       & $Y=0$                  & $P(D=1)$          & $P(D=1|Y=0, A=a)=P(D=1|Y=0, A \neq a)$             \\ \hline
\multirow{2}{*}{\textbf{Sufficiency}} & Parity of positive predictive values & $D=1$                  & $P(Y=1)$          & $P(Y=1|D=1, A=a)=P(Y=1|D=1, A \neq a)$             \\ \cline{2-5} 
                                      & Parity of false omission rates       & $D=0$                  & $P(Y=1)$          & $P(Y=1|D=0, A=a)=P(Y=1|D=0, A \neq a)$             \\ \hline
\end{tabular}
\end{table*}
\section{Justifications for group fairness measures}\label{sec:philosophy}

Creating a fair decision system means enforcing one or several fairness criteria by either manipulating the training data, the prediction model or the decision rule.
Note that enforcing \emph{all} fairness metrics at the same time is a mathematical impossibility in most cases (and thus essentially always in practice) \cite{fairmlbook, kleinberg2016inherent, chouldechova2017fair}.
Therefore, we usually have to make a choice between these fairness metrics. This choice needs a justification, and this is a moral, not a technical, justification. In the following we will discuss some justifications for the different fairness criteria. Since the application context is highly relevant for which justification is deemed the most fitting one, it is not surprising that, in different contexts, different fairness metrics might be suggested. Also, people might differ in their views on which justification is valid. However, the goal of this section is to make clear which justifications are coupled with the different fairness criteria.

Note that we focus on fairness \textit{with respect to the decision subjects}, i.e., the people directly affected by the decision.
What is considered to be morally adequate for decision-makers or wider society is a different discussion.
One might, for example, find that statistical parity would be fair towards the decision subjects, but would mean a loss in performance of the \gls{ADM} system, which in turn would lead to an indefensible loss in profit to the decision-maker.
In this case, the moral assessment of fairness metrics might differ for different stakeholders. However, we will focus on what it means to be fair towards the decision subjects.
This discussion is by no means meant to be exhaustive, since the philosophical literature on this topic is still very much a work in progress.

\paragraph{Statistical parity}

Several works (see, e.g., \cite{binns2020apparent, raez2021group, friedler2016possibility, hertweck2021moral}) argue that if differences in predictive features $\vec{X}$ are due to unjust historical injustices or unjust societal structures (such as male students being more often encouraged to go into STEM fields than their equally talented female counterparts), then these differences should be corrected for by enforcing statistical parity.
\cite{hertweck2021moral} argues that past and current injustices alone are, however, not sufficient for justifying enforcing statistical parity: It is also crucial to consider the effects of the intervention.
For a case of medical treatment, they show that enforcing statistical parity might actually harm the already underprivileged group.

Therefore, enforcing statistical parity might make sense if (1) there are differences in $\vec{X}$ that are caused by unjust circumstances and that would therefore, without intervention, lead to unjustifiable and unjust differences in $D$, and if (2) this intervention at least does not harm the already underprivileged group.

\paragraph{Conditional statistical parity}

As a relaxation of statistical parity, conditional statistical parity also assumes that fairness means equality in the (average) decision for the two groups.
The difference is that equality of decisions should only hold when controlling for some elements of $\vec X$, called "legitimate" attributes \cite{corbett2017algorithmic, vzliobaite2011handling}.
Focusing on \emph{conditional} statistical parity is thus motivated by the assertion that differences in certain attributes justify differences in the decision $D$: statistical parity should hold, but only for people with, e.g., a stellar credit history.
However, the choice of legitimate attributes has to be well-justified.
When we, for example, argue that differences in the credit history justify differences in treatment, we might reason that people are fully responsible for their credit history -- which is an assumption that does not necessarily hold.
To see this, let us consider an example provided in \cite{binns2020apparent}: Assume that bank clerks are more lenient with the repayment deadlines of men compared to women.
In this case, the credit history is unfit as a legitimate attribute as it is biased.
In fact, if we compared the credit repayment rates of women and men with similar credit histories, we would find that women repay their loans at a higher rate than men.
One may thus argue that the credit approval rate of women in this subpopulation of individuals with similar credit scores should be higher than that of men.
Since conditional statistical parity demands that female and male applicants \textit{with the same credit history} are accepted at the same rate, the metric would (in this case) disadvantage female applicants.

We conclude that conditional statistical parity is appropriate when the "equal treatment" paradigm can justifiably be applied to subpopulations defined by legitimate attributes (rather than to the full population).

\paragraph{Separation}

Separation is fundamentally different from independence in that the "equal treatment" paradigm is conditioned on $Y$: Equal treatment should hold for individuals with the same value of $Y$, but not necessarily when comparing individuals with different $Y$.
For example, one may argue that it is not necessary that a credit lending system treats women and men, on average, equally.
Instead, it might be sufficient if women and men, who repay their loan, are treated equally and if women and men, who default on their loan, are also treated equally.
Thus, separation explicitly accepts treatment differences based on $Y$ and assume that they do not constitute a source of unfairness per se.
This is, for example, the case when $D$ is about granting a benefit or access to resources, such as college admissions or job promotions, where $Y=1$ constitutes the main reason for $D=1$, and this is morally right.
Note that a necessary condition for focusing on separation is that $Y$ is not biased with respect to $A$ (as it would, for example, be the case if the bank was more lenient with the credit repayment deadlines of men compared to women, so that $Y$ does not actually represent timely repayments, but a biased view of it).
\cite{loi2021fair} therefore argues that separation is appropriate to use if $Y$ corresponds to what \emph{morally} justifies a beneficial or harmful outcome.
Sometimes, the moral justification is given only if the individuals can be made \emph{responsible} for $Y$.
However, this is not always the case.
Sometimes, treating people differently based on $Y$ is morally justified even if they are not responsible for their $Y$.
For example, in health-care, it may be morally legitimate to decide on giving a therapy to people based on whether or not they will be cured through the therapy (where whether or not they get cured corresponds to $Y$), irrespective of whether they are responsible for their current health issues; in credit lending it may be considered morally legitimate to deny a loan to someone who cannot repay it, even if the individual is not morally responsible for the failure to repay, etc.

Separation is therefore justified if we believe that inequalities in treatment between groups are acceptable as long as people with the same $Y$ are treated equally.

\paragraph{Sufficiency}

Focusing on sufficiency means looking at the subpopulations defined by $D$ and assessing them with respect to $Y$.
The decision system is considered fair if the average propensity for $Y=1$ is equal for both groups.
For the credit lending example, we assess whether the accepted men have the same repayment rate as the accepted women and whether the rejected men have the same repayment rate as the rejected women.
We first of all note that experimental evidence has shown that sufficiency is oftentimes (close to) fulfilled in unconstrained learning (i.e., whenever there is no specific intervention to achieve a certain kind of fairness) \cite{liu2019implicit}.

Let us now think about what assumptions would morally justify enforcing sufficiency.
For this, we consider the motivation behind a relaxation of sufficiency which is often used in practice, e.g., in measuring bias in policing \cite{simoiu2017problem}: the "outcome test" (which aligns with testing for equality in positive predictive values (PPVs)).
The test was proposed by the American economist Gary Becker \cite{becker2010economics} to measure discrimination in credit lending.
As an example, assume that a bank estimates each credit loan applicant's repayment probability and gives a loan to men if they have a repayment probability above 30\%, but to women only if their repayment probability is above 70\%.
In this case, it seems clear that women, who receive a loan, repay it at a much higher rate than men.
Since in real life the thresholds above which the bank grants loans to women and men are not known, the observation of different repayment rates might be used to infer a systematic difference in how women and men are treated.
However, differences in repayment rates do not mathematically guarantee differences in thresholds because many different combinations of \textit{individuals who apply for a loan} and \textit{thresholds} could lead to the same results in the outcome test.
This is what is known as the \textit{infra-marginality problem}.\footnote{
For more detailed information on this problem and a way to infer the thresholds from the data, see \cite{simoiu2017problem}.
}
Therefore, the intuition that different threshold have been applied when sufficiency is violated does not hold.
\cite{simoiu2017problem} shows that this is not only a theoretical issue, but also a practical one.
Since sufficiency does not measure differences in thresholds, it is still unclear why sufficiency would be morally relevant in the fairness discussion.

A possible argument for focusing on sufficiency is given in \cite{loi2021fair}. 
According to the generalized theory in \cite{loi2021fair}, sufficiency is appropriate when a different decision $D$ justifies different expectations for the individuals of obtaining a given (beneficial or harmful) outcome $Y$.
This might be plausible when the decision $D$ corresponds to the decision to give a recommendation or warning message such as "it is dangerous \textit{for you} to rent this Ferrari".
In this case, it is morally appropriate for people to have different expectations of getting hurt in driving the car depending on whether they got this warning message or not.
People who receive the message, for example, rightly expect that they are more likely to get hurt than people who do not receive the message.
This affects the kind of moral complaint they may have against the rental company if they still decide to drive the car.

Sufficiency therefore appears to be appropriate if we believe that the individuals about whom decisions are made justly have different expectations of their $Y$ depending on their $D$.
\section{Conclusion}

We provided an introduction to group fairness for data scientists by considering how different socio-demographic groups may be compared in the context of prediction-based decision systems.
In particular, we focused on the question of what parts of the full population are pertinent for assessing the question of fairness.
Starting from the formal definition of a prediction-based decision problem with the variables $A,\vec X,Y,D$, we described six fairness metrics that cover different perspectives of how to compare socio-demographic groups with respect to decision fairness.
In line with existing literature, we pointed to three broad categories of group fairness under which these metrics fall: independence, separation and sufficiency.
This approach to group fairness can help practitioners understand the choices they have if they want to enforce a group fairness metric.
Since the literature on philosophical guidance is not evolved enough yet to provide concise rules for when to apply each criterion, we provided some pointers to aspects that have to be considered when choosing a fairness metric.

\section*{Acknowledgment}

This work was supported by the National Research Programme “Digital Transformation” (NRP 77) of the Swiss National Science Foundation (SNSF), grant number 187473.
We thank Michele Loi as well as our anonymous reviewers for their helpful feedback.

\bibliographystyle{IEEEtran}
\bibliography{main}

\end{document}